\documentclass{article}
\usepackage{spconf,amsmath,graphicx}
\usepackage{hyperref}
\usepackage{amsfonts,bm, dsfont, amssymb, amsthm, verbatim}
\usepackage{enumitem}
\setitemize{noitemsep,topsep=2pt,parsep=0pt,partopsep=2pt}

\newcommand{\ma}[1]{\ensuremath{\mathsf{#1}}}
\renewcommand{\vec}[1]{\ensuremath{\mathbf{#1}}}
\newtheorem{proposition}{Proposition}
\newcommand{\ie}{\textit{i.e.}}
\DeclareMathOperator*{\argmin}{\mathrm{argmin}}

\title{SMOOTHING GRAPH SIGNALS VIA RANDOM SPANNING FORESTS}
	\name{Yusuf Y. Pilavci, Pierre-Olivier Amblard, Simon Barthelm\'e, Nicolas Tremblay\thanks{This work was partly funded by the ANR GenGP (ANR-16-CE23-0008), the ANR GraVa (ANR-18-CE40-0005), the LabEx PERSYVAL-Lab (ANR-11-LABX-0025-01), the CNRS PEPS I3A (Project RW4SPEC), the Grenoble Data Institute (ANR-15-IDEX-02), the LIA CNRS/Melbourne Univ Geodesic, and the MIAI chair "LargeDATA
			 at UGA." }}
\address{CNRS, Univ. Grenoble Alpes, Grenoble INP, GIPSA-lab, Grenoble, France}
\begin{document}
%
\maketitle
\begin{abstract}
Another facet of the elegant link between random processes on graphs and Laplacian-based numerical linear algebra is uncovered: based on random spanning forests, novel Monte-Carlo estimators for graph signal smoothing are proposed. These random forests are sampled efficiently via a variant of Wilson's algorithm --in time linear in the number of edges. The theoretical variance of the proposed estimators are analyzed, and their application to several problems are considered, such as Tikhonov denoising of graph signals or semi-supervised learning for node classification on graphs.
\end{abstract}
\section{Introduction}
\label{sec:intro}
Tikhonov denoising of graph signals~\cite{shuman_emerging_2013, sandryhaila_big_2014}, semi-supervised learning for node classification in graphs~\cite{avrachenkov_generalized_nodate}, post-sampling graph signal reconstruction~\cite{puy_random_2016} are a few examples falling in the following class of optimization problems. 
Given a graph signal $\vec{y}=\left(y_1|\ldots|y_n\right)^t\in\mathbb{R}^n$ where $y_i$ is the value measured at node $i$ of a graph, one considers the problem
\vspace{-0.1cm}\begin{equation}
\label{eq:linear_estimation}
\hat{\vec{x}} = \argmin_{\vec{z}\in\mathbb{R}^n} q \left|\left| \vec{y} - \vec{z} \right| \right|^{2} + \vec{z}^{t} \ma{L} \vec{z},\vspace{-0.15cm}
\end{equation}
where $\ma{L}$ is the Laplacian of the graph and $q>0$ a parameter tuning the trade-off between a data-fidelity term $\left|\left| \vec{y} - \vec{z} \right| \right|^{2}$ --encouraging the solution $\hat{\vec{x}}$ to be close to $\vec{y}$-- and a smoothness term $\vec{z}^{t} \ma{L} \vec{z}$ --encouraging the solution $\hat{\vec{x}}$ to vary slowly along any path of the graph. In Tikhonov denoising, $\vec{y}$ is a noisy measurement of an underlying signal $\vec{x}$ that one wants to recover. In semi-supervised learning (SSL), $\vec{y}$ are known labels that one wants to propagate on the graph to classify all the nodes in different classes (see Section~\ref{sec:experiments} for more details). 

This optimization problem admits an explicit solution:
\vspace{-0.08cm}
\begin{equation}
\label{eq:linear_estimation_sol}
\hat{\vec{x}} = \ma{K} \vec{y} \quad\text{with}\quad \ma{K}=q(q\ma{I}+\ma{L})^{-1}\in\mathbb{R}^{n\times n},
\end{equation}
that requires the inversion of a regularized Laplacian $q\ma{I}+\ma{L}$, where $\ma{I}$ is the identity matrix. Computing $\ma{K}$ costs $\mathcal{O}(n^3)$ elementary operations and becomes prohibitive as $n$ increases. \\

\vspace{-0.0cm}\noindent\textbf{State-of-the-art.} For large $n$ (say $\geq 10^4$), the state-of-the-art includes iterative methods such as conjugate gradient with preconditioning~\cite{saad_iterative_2003} where $\ma{L}$ is only accessed via matrix-vector multiplications of the form $\ma{L}\vec{z}$, and polynomial approximation methods~\cite{shuman_chebyshev_2011} that approximate $\ma{K}$ by a low-order polynomial in $\ma{L}$. Both classes of methods  enable to compute $\hat{\vec{x}}$ in time linear in $|\mathcal{E}|$, the number of edges of the graph.\\

\vspace{-0.2cm}\noindent\textbf{Contribution.} We introduce two Monte-Carlo estimators of $\hat{\vec{x}}$ based on random spanning forests, that:
\begin{itemize}
	\item show another facet of the elegant link between random processes on graphs and Laplacian-based numerical linear algebra, such as in~\cite{chung_spectral_1997,agaev_spanning_2001,barthelme_estimating_2019}
	\item scale linearly in $|\mathcal{E}|$ and thus useful on very large graphs 
	\item can be implemented in a fully distributed fashion: as long as each node is able to communicate with its neighbours, the result can be obtained  without centralized knowledge of the graph's structure $\ma{L}$ (see the implementation paragraph at the end of Section~\ref{sec:estimators}). 
\end{itemize}
The Julia code implementing these estimators and reproducing this papers' results is available on the authors' website\footnote{\href{http://www.gipsa-lab.fr/~nicolas.tremblay/files/graph_smoothing_via_RSF.zip}{www.gipsa-lab.fr/${\raise.15ex\hbox{$\scriptstyle\sim$}}$nicolas.tremblay/files/graph$\_$smoothing$\_$via$\_$RSF.zip}}. \\

\vspace{-0.2cm}\noindent{\textbf{Structure of the paper.}} We provide the necessary notations and background in Section~\ref{sec:background}. In Section~\ref{sec:estimators}, we detail the two novel estimators before generalizing them to cases where $\ma{K}$ is of the form $\ma{K}=(\ma{Q}+\ma{L})^{-1}\ma{Q}$ with $\ma{Q}$ a positive diagonal matrix. The experimental Section~\ref{sec:experiments} gives an illustration on image denoising before showing how to use our estimators in the context of SSL. We conclude in Section~\ref{sec:conclusion}.

\section{Background}
\label{sec:background}

\vspace{-0.1cm}\noindent\textbf{Notations and preliminary definitions.} We consider undirected graphs $\mathcal{G}=(\mathcal{V}, \mathcal{E}, \ma{W})$ where $\mathcal{V}$ is the set of $|\mathcal{V}|=n$ nodes, $\mathcal{E}$ the set of $|\mathcal{E}|$ edges, and $\ma{W}\in\mathbb{R}^{n\times n}$ the symmetric weighted adjacency matrix. 
We denote by $d_i=\sum_j \ma{W}_{ij}$ the degree of node $i$, $\vec{d}=(d_1,d_2,\ldots,d_n)^t\in\mathbb{R}^{n}$ the degree vector,  and $\ma{D}=\text{diag}(\vec{d})\in\mathbb{R}^{n\times n}$ the diagonal degree matrix. In this paper, we consider the Laplacian matrix defined as $\ma{L}=\ma{D}-\ma{W}\in\mathbb{R}^{n\times n}$. $\ma{L}$ is positive semi-definite (PSD)~\cite{chung_spectral_1997} and its eigenvalues can be ordered as $0=\lambda_1\leq\lambda_2\leq\ldots\leq\lambda_n$. 
We also consider graph signals $\vec{x}\in\mathbb{R}^n$ defined over the nodes $\mathcal{V}$: $x_i=x(i)$ is the signal's value at node $i$. We denote by $\vec{1}\in\mathbb{R}^n$ the constant vector equal to 1, and by $\bm{\delta}_i\in\mathbb{R}^n$ the Kronecker delta such that $\delta_i(j)=1$ if $j=i$, and $0$ otherwise.

A tree of $\mathcal{G}$ is understood as a connected subgraph of $\mathcal{G}$ that does not contain cycles. A rooted tree of $\mathcal{G}$, \ie, a tree of $\mathcal{G}$ whose edges are all oriented towards one node called root, is generically denoted by $\tau$. A rooted forest of $\mathcal{G}$, \ie, a set of disjoint rooted trees, is generically denoted by $\phi$. In the following, we only consider rooted trees and rooted forests, and thus simply refer to them as trees and forests. Also, $\rho(\phi)$ stands for the set of roots of the trees in $\phi$. As such, $|\rho(\phi)|$ is the total number of trees in $\phi$. We define the function $r_\phi:\mathcal{V}\rightarrow\mathcal{V}$ such that $r_\phi(i)$ is the root of the tree containing node $i$ in the forest $\phi$. Node $i$ is said to be rooted in $r_\phi(i)$. \\

\vspace{-0.2cm}\noindent\textbf{Random spanning forests (RSFs).} Let us recall that a spanning tree (resp. forest) of $\mathcal{G}$ is a tree (resp. forest) of $\mathcal{G}$ that spans all nodes in $\mathcal{V}$.
Now, for a given graph $\mathcal{G}$, there exists in general many spanning trees (and even more spanning forests). 
Probabilists and computer scientists have been studying several distributions over spanning trees and forests in the past. A classical distribution over spanning trees is:
\begin{align}
\mathbb{P}(T=\tau) \propto \prod_{(ij)\in\tau} \ma{W}_{ij}. 
\end{align}
Sampling from this distribution yields a so-called uniform\footnote{this terminology comes from the fact that in unweighted graphs, this distribution reduces to the uniform distribution over all spanning trees} spanning tree (UST) $T$, and can be efficiently performed by Wilson's algorithm~\cite{wilson_generating_1996} via loop-erased random walks.

We focus here on the following parametrized distribution over spanning forests:
\begin{align}
\label{eq:distrib_forest_q}
\mathbb{P}(\Phi_q=\phi)\propto q^{|\rho(\phi)|}\prod_{\tau\in\phi}\prod_{(ij)\in\tau} \ma{W}_{ij},\quad q\in\mathbb{R}^{+*}.
\end{align}
Sampling from this distribution yields a RSF  $\Phi_q$ and is efficiently performed (in time $\mathcal{O}(|\mathcal{E}|/q)$) via a variant of Wilson's algorithm~\cite{avena_two_2017}. 
Many properties of $\Phi_q$ are known~\cite{avena_two_2017, avena_random_2013}. For instance, the probability that node $i$ is rooted in $j$ is (see lemma 3.3 in~\cite{avena_random_2013})
	\begin{align}
		\label{eq:Kij}
		\mathbb{P}\left(r_{\Phi_q}(i)=j\right)=\ma{K}_{ij}.
	\end{align}


\section{RSF-based estimators}
\label{sec:estimators}
\vspace{-0.1cm}
Given a signal $\vec{y}$, our goal is to compute $\hat{\vec{x}}=\ma{K}\vec{y}$ without computing explicitly $\ma{K}$, that is, without inverting $q\ma{I}+\ma{L}$. We define two Monte-Carlo estimators of $\hat{\vec{x}}$, both based on RSFs. \\

\vspace{-0.15cm}\noindent\textbf{A first estimator of $\hat{\vec{x}}$.} Let us consider a realisation $\Phi_q$ of RSF and define the estimator $\tilde{\vec{x}}$ as
\begin{align}
\label{eq:tildex}
\forall i\quad\tilde{x}(i) = y\left[r_{\Phi_q}(i)\right].
\end{align}
\begin{proposition}
$\tilde{\vec{x}}$ is unbiased: $\mathbb{E}\left[\tilde{\vec{x}}\right]=\hat{\vec{x}}$. Moreover:
$$\mathbb{E}\left(||\tilde{\vec{x}}-\hat{\vec{x}}||^2\right)=\sum_i \text{\emph{var}}(\tilde{x}(i)) = \vec{y}^t(\ma{I}-\ma{K}^2)\vec{y}.$$ 
\end{proposition}
\vspace{-0.0cm}
\begin{proof}
	Let $i\in\mathcal{V}$. One has, using Eq.~\eqref{eq:Kij}:
	\begin{align}
	\mathbb{E}\left[\tilde{x}(i)\right]&=\mathbb{E}\left[y\left[r_{\Phi_q}(i)\right]\right]=\sum_{j} \mathbb{P}(r_{\Phi_q}(i)=j) y_j\\
	&=\sum_{j} \ma{K}_{ij}y_j=\bm{\delta}_i^t\ma{K}\vec{y}=\hat{x}(i).
	\end{align}
	$\tilde{\vec{x}}$ is thus unbiased. A similar calculation yields  $\mathbb{E}\left[\tilde{x}(i)^2\right]=\bm{\delta}_i^t\ma{K}\vec{y}^{(2)}$ where $\forall k, y^{(2)}_k=y_k^2$,  such that the variance verifies:
	$$\text{var}(\tilde{x}(i)) = \mathbb{E}\left[\tilde{x}(i)^2\right] - \mathbb{E}\left[\tilde{x}(i)\right]^2=\bm{\delta}_i^t\ma{K}\vec{y}^{(2)}-(\bm{\delta}_i^t\ma{K}\vec{y})^2.$$
	Summing over all nodes yields:
	$$\sum_i \text{var}(\tilde{x}(i))=\vec{1}^t\ma{K}\vec{y}^{(2)}-\vec{y}^t\ma{K}^2\vec{y}.$$
	Noticing that $\vec{1}$ is an eigenvector of $\ma{K}$ with eigenvalue $1$ (as $\vec{1}$ is an eigenvector of $\ma{L}$ with eigenvalue $0$) ends the proof.
\end{proof}

\vspace{-0.0cm}\noindent\textbf{An improved estimator of $\hat{\vec{x}}$.} A random forest $\Phi_q$ contains $|\rho(\Phi_q)|$ trees that we enumerate from $1$ to $|\rho(\Phi_q)|$. Consider  $\mathcal{V}_k\subset\mathcal{V}$  the subset of nodes that are in the $k$-th tree. As $\Phi_q$ is a spanning forest,  
$\mathcal{P}=\{\mathcal{V}_1,\mathcal{V}_2,\ldots,\mathcal{V}_{|\rho(\Phi_q)|}\}$ is a partition of $\mathcal{V}$ (\ie, a set of disjoint subsets that cover $\mathcal{V}$). Let $t$ be the tree membership function that associates to each node $i$ the tree number $t(i)$ to which it belongs. For instance, $|\mathcal{V}_{t(i)}|$ is the size of the tree containing node $i$. We define a second estimator as
\begin{align}
\forall i\quad\bar{x}(i) = \frac{1}{|\mathcal{V}_{t(i)}|}\sum_{j\in\mathcal{V}_{t(i)}} y_j.
\end{align}
\vspace{-.0cm}\begin{proposition}
	$\bar{\vec{x}}$ is unbiased: $\mathbb{E}\left[\bar{\vec{x}}\right]=\hat{\vec{x}}$. Moreover:
	$$\mathbb{E}\left(||\bar{\vec{x}}-\hat{\vec{x}}||^2\right)=\sum_i \text{\emph{var}}(\bar{x}(i)) = \vec{y}^t(\ma{K}-\ma{K}^2)\vec{y}.$$ 
\end{proposition}
\vspace{-.0cm}\begin{proof}
	Let us denote by $\pi$ the function that takes as entry a forest $\phi$ and outputs its associated partition. Let us also define $s_{ij}=1$ if $t(i)=t(j)$, and $0$ otherwise. 
	We need the Proposition 2.3 of~\cite{avena_two_2017}. Fixing a partition $\mathcal{P}$ of $\mathcal{V}$, one has: $$\mathbb{P}\left(r_{\Phi_q}(i)=j|\pi(\Phi_q)=\mathcal{P}\right)=\frac{s_{ij}}{|\mathcal{V}_{t(i)}|}.$$
	In words, this means that given a partition, the distribution of the root within each part $\mathcal{V}_l$ is uniform over $\mathcal{V}_l$. 
	Also, note that $\ma{K}_{ij}=\mathbb{P}\left(r_\phi(i)=j\right)$ can be written:
	\begin{align}
	 \ma{K}_{ij}&=\sum_\mathcal{P}\mathbb{P}\left(r_{\Phi_q}(i)=j|\pi(\Phi_q)=\mathcal{P}\right)\mathbb{P}\left(\pi(\Phi_q)=\mathcal{P}\right)\nonumber\\
	 &=\sum_\mathcal{P} \frac{s_{ij}}{|\mathcal{V}_{t(i)}|} \mathbb{P}\left(\pi(\Phi_q)=\mathcal{P}\right)\nonumber\\
	 &=\sum_\mathcal{P} \frac{s_{ij}}{|\mathcal{V}_{t(i)}|} \sum_{\phi\text{ s.t. } \pi(\phi)=\mathcal{P}}\mathbb{P}\left(\Phi_q=\phi\right)\nonumber\\
	 &=\sum_\phi \mathbb{P}\left(\Phi_q=\phi\right) \frac{s_{ij}}{|\mathcal{V}_{t(i)}|} = \mathbb{E}\left(\frac{s_{ij}}{|\mathcal{V}_{t(i)}|}\right).\label{eq:last}
	\end{align}
	Given a RSF $\Phi_q$, define the matrix $\ma{S}_{ij}= \frac{s_{ij}}{|\mathcal{V}_{t(i)}|}\in\mathbb{R}^{n\times n}$. Eq.~\eqref{eq:last} translates as: $\mathbb{E}(\ma{S})=\ma{K}$. Thus, $\forall i\in\mathcal{V}$:
	\begin{align}
	\mathbb{E}\left[\bar{x}(i)\right]&=\mathbb{E}\left[\frac{1}{|\mathcal{V}_{t(i)}|}\sum_{j\in\mathcal{V}_{t(i)}} y_j\right]\\
	&=\mathbb{E}\left[\bm{\delta}_i^t\ma{S}\vec{y}\right]=\bm{\delta}_i^t\mathbb{E}\left[\ma{S}\right]\vec{y}=\bm{\delta}_i^t\ma{K}\vec{y}=\hat{\vec{x}}_i.
	\end{align}   
	The estimator $\bar{\vec{x}}$ is thus unbiased. 
	Note that one also has:
	\begin{align}
	\mathbb{E}\left[\bar{x}(i)^2\right] = 	\mathbb{E}\left[\left(\bm{\delta}_i^t \ma{S}\vec{y}\right)^2\right] = \vec{y}^t\mathbb{E}\left[\ma{S}^t\bm{\delta}_i\bm{\delta}_i^t \ma{S}\right]\vec{y},
	\end{align}
	such that
	\begin{align}
	\sum_i \text{var}(\bar{x}(i))&= \sum_i \vec{y}^t\mathbb{E}\left[\ma{S}^t\bm{\delta}_i\bm{\delta}_i^t \ma{S}\right]\vec{y} -(\bm{\delta}_i^t\ma{K}\vec{y})^2\\
	&=\vec{y}^t\left[\mathbb{E}\left[\ma{S}^t\ma{S}\right]-\ma{K}^2\right]\vec{y}.
	\end{align}
	Note that $\ma{S}^t\ma{S}=\ma{S}$, \ie, $\mathbb{E}\left[\ma{S}^t\ma{S}\right]=\ma{K}$, finishing the proof.
\end{proof}

\vspace{-0.cm}\noindent\textbf{Rao-Blackwellisation.} Note that $\bar{\vec{x}}$
is a Rao-Blackwellisation of $\tilde{\vec{x}}$ where the sufficient statistic is
the partition induced by the RSF. As such, $\bar{\vec{x}}$ is necessarily an
improvement over $\tilde{\vec{x}}$. This improvement can also be observed in the
variance equations of both propositions: as $\ma{K}$ is PSD with eigenvalues
between $0$ and $1$, one has: $\forall\vec{y}; \ \vec{y}^t(\ma{K}-\ma{K}^2)\vec{y}\leq\vec{y}^t(\ma{I}-\ma{K}^2)\vec{y}$.\\

\vspace{-0.2cm}\noindent\textbf{Tikhonov denoising.} Let $\vec{y}=\vec{x}+\bm{\epsilon}$\; be a noisy measurement of a signal $\vec{x}$ that one wishes to recover. Assuming that the measurement noise $\bm{\epsilon}$ is Gaussian with covariance  $\mathbb{E}_{\bm{\epsilon}} \left(\bm{\epsilon}\bm{\epsilon}^t\right)=\gamma^2\ma{I}$, one can write (for instance for the second estimator):
\begin{align}
\mathbb{E}_{\bm{\epsilon}}\left[\mathbb{E}\left(||\bar{\vec{x}}-\hat{\vec{x}}||^2\right)\right] &=  \vec{x}^t(\ma{K}-\ma{K}^2)\vec{x} + \gamma^2 \text{Tr}\left(\ma{K}-\ma{K}^2\right)\nonumber\\
&=||\bar{\ma{F}}\vec{x}||_2^2 + \gamma^2 \text{Tr}\left(\bar{\ma{F}}^2\right),\nonumber
\end{align}
where $\bar{\ma{F}}=\ma{U}\bar{f}(\ma{\Lambda})\ma{U}^t$ is a graph bandpass filter~\cite{shuman_emerging_2013,djuric_chapter_2018} with frequency response  $\bar{f}(\lambda)=\frac{\sqrt{q\lambda}}{q+\lambda}$. The second term depends on the noise level $\gamma$. The first term, however, depends on the original signal $\vec{x}$ filtered by $\bar{f}$: the Fourier components of $\vec{x}$ associated with graph frequencies around $\lambda=q$ (maximizing $\bar{f}$) are thus harder to denoise than the ones close to $0$ or $\lambda_n$.\\ 

\vspace{-0.2cm}\noindent\textbf{In practice.} For a given $q$, $N$ independent RSFs are sampled --in time $\mathcal{O}(\frac{N|\mathcal{E}|}{q})$. Each RSF provides an independent estimate of $\hat{\vec{x}}$, and all $N$ estimates are finally averaged. \\

\vspace{-0.2cm}\noindent\textbf{Generalisation.} Instead of estimating results of the form $\left(q\ma{I}+\ma{L}\right)^{-1}q\vec{y}$, one may need to estimate results of the form $\left(\ma{Q}+\ma{L}\right)^{-1}\ma{Q}\vec{y}$ where $\ma{Q}=\text{diag}(\vec{q})$ is a diagonal matrix, with $\vec{q}=\left(q_1|\ldots|q_n\right)^t\in\;(0,+\infty)^n$. In order to tackle this case, one considers the following distribution over forests:
\begin{align}
\label{eq:distribt_RSF}
\mathbb{P}(\Phi_\ma{Q}=\phi)\propto \prod_{r\in\rho(\phi)} q_r\prod_{\tau\in\phi}\prod_{(ij)\in\tau} \ma{W}_{ij},
\end{align}
that generalizes the distribution of  Eq.~\eqref{eq:distrib_forest_q}. One can efficiently sample from this distribution --also via a variant of Wilson's algorithm (see the next paragraph). The introduced estimators generalize naturally to this case. In fact, given a RSF $\Phi$, their formulation is exactly the same, the sole difference stemming from the distribution from which $\Phi$ is drawn from. Propositions 1 and 2 are still correct (proofs are omitted due to lack of space) for $\ma{K}=\left(\ma{Q}+L\right)^{-1}\ma{Q}$ instead of $\ma{K}=\left(q\ma{I}+L\right)^{-1}q\ma{I}$.\\

\vspace{-0.2cm}\noindent\textbf{Implementation.} In a nutshell, an algorithm sampling from the distribution of  Eq.~\eqref{eq:distribt_RSF} i/~adds a node $\Delta$ to the graph and connects it to each node $i$ with weight $q_i$; ii/~runs Wilson's \texttt{RandomTreeWithRoot} algorithm (based on loop-erased random walks --see Figure 1 of~\cite{wilson_generating_1996}) on this augmented graph to sample a UST $T$ rooted in $\Delta$; iii/~cuts the edges connected to $\Delta$ in $T$ yielding a RSF $\Phi_\ma{Q}$. Then, the estimator $\tilde{\vec{x}}$ associates to each node $i$ the value of $\vec{y}$ at its root $r_{\Phi_\ma{Q}}(i)$, whereas the estimator $\bar{\vec{x}}$ associates to each node $i$ the average of $\vec{y}$ over all the nodes belonging to its tree. All these operations can be done in a distributed fashion: no centralized knowledge of the graph is needed. Also, once the RSF is sampled, the computations involved for the estimators are not only distributed but can also be made in parallel (within each tree of the forest). To give an order of magnitude of computation times, for a random vector $\vec{y}$ and $\vec{q}=\vec{1}$, our naive Julia implementation of $\bar{\vec{x}}$ on a graph with $n=10^5$ (resp. $10^6$) nodes and $|\mathcal{E}|=10^6$ (resp $10^7$) edges runs in average in 35 (resp. 550) ms on a laptop. These times are to compare to the optimized built-in sparse matrix multiplication $\ma{L}\vec{y}$ running in 6 (resp. 115) ms, which is the building block of both conjugate gradient and polynomial approximation methods stated in the introduction. Our methods are thus comparable to the state-of-the-art in computation time.


\begin{figure*}[t]
	\centering
	\begin{minipage}{0.245\linewidth}
		\begin{center}
			$\vec{x}$:\\
			\includegraphics[width=.7\columnwidth]{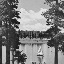}\\\vspace{0.1cm}
			
			$\tilde{\vec{x}} (N=1)$:\\
			\includegraphics[width=.7\columnwidth]{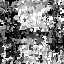}\\
		\end{center}
	\end{minipage}
	\hfill
	\begin{minipage}{0.245\linewidth}
		\begin{center}
			$\vec{y}$:\\
			\includegraphics[width=.7\columnwidth]{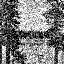}\\\vspace{0.1cm}
			
			$\bar{\vec{x}} (N=1)$:\\
			\includegraphics[width=.7\columnwidth]{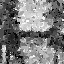}\\
		\end{center}
	\end{minipage}
	\hfill
	\begin{minipage}{0.245\linewidth}
		\begin{center}
			$\hat{\vec{x}}$:\\
			\includegraphics[width=.7\columnwidth]{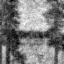}\\\vspace{0.1cm}
			
			$\tilde{\vec{x}} (N=20)$:\\
			\includegraphics[width=.7\columnwidth]{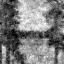}\\
		\end{center}
	\end{minipage}
	\hfill
	\begin{minipage}{0.245\linewidth}
		\begin{center}
			\vspace{.4cm}\includegraphics[width=\columnwidth]{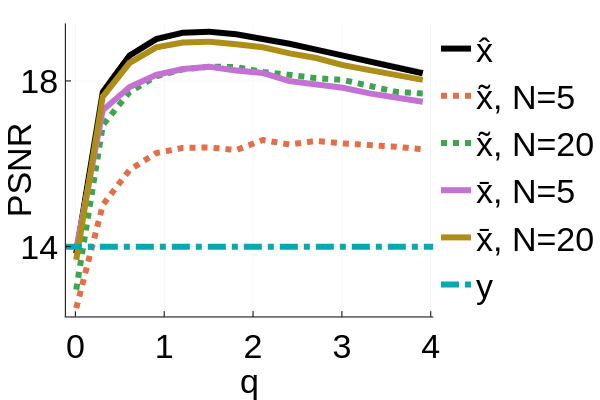}\\\vspace{.2cm}
			
			$\bar{\vec{x}} (N=20)$:\\
			\includegraphics[width=.7\columnwidth]{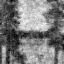}\\
		\end{center}
	\end{minipage}
	
	\caption{Illustration on an image. A grayscale image $\vec{x}$ is considered as a graph signal on an unweighted 2D grid graph where each pixel is a node connected to its four immediate neighbours. $\vec{y}=\vec{x}+\bm{\epsilon}$ is a noisy measurement of $\vec{x}$ ($\bm{\epsilon}$ is Gaussian with covariance matrix $\gamma^2\ma{I}$). $\hat{\vec{x}}=q\left(q\ma{I}+\ma{L}\right)^{-1}\vec{y}$ is the exact Tikhonov denoised signal (here with $q=1$) that we try to estimate. Bottom line: the two left images show estimates of $\hat{\vec{x}}$ obtained with the RSF-based estimators $\tilde{\vec{x}}$ and $\bar{\vec{x}}$ detailed in Section~\ref{sec:estimators}. Averaging over $N=20$ forest realisations, one obtains the two images on the right. Finally, the top-right figure is the Peak Signal-to-Noise Ratio (PSNR) of the denoised images (averaged over $100$ realisations of $\bm{\epsilon})$. As usual in these scenarios, there exists an optimal regularization parameter (a value of $q$ maximizing the PSNR) that is here observed to be between $1$ and $2$. }
	\vspace{-0.3cm}
	\label{fig:images}
\end{figure*}

\begin{figure}
	\centering
	\includegraphics[width=.49\columnwidth]{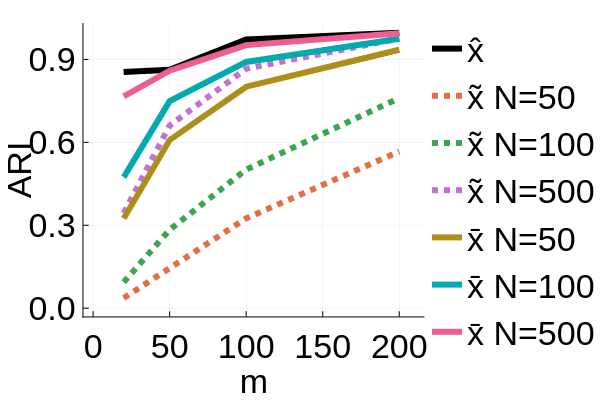}\hfill
	\includegraphics[width=.49\columnwidth]{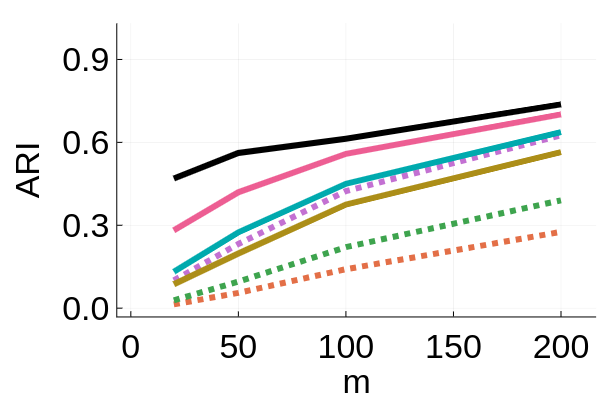}
	\caption{Illustration on SSL. Performance of the community recovery in a SBM with two equal-size classes, vs. the number of pre-labeled nodes $m$ in each class. Results are averaged over 10 realisations of SBM. Left: setting with strong communities. Right: setting with fuzzy communities.}
	\vspace{-0.3cm}
	\label{fig:SSL_exp}
\end{figure}

\section{Experiments}
\label{sec:experiments}
\vspace{-0.1cm}\noindent{\textbf{Illustration on an image.}} Fig.~\ref{fig:images} shows an image denoising example on a $64\times 64$ grayscale image.  Constant irregular patches are observed on the realisation of $\tilde{\vec{x}} (N=1)$: they are the trees of the associated RSF realisation.  Also, as expected, $\bar{\vec{x}}$ converges faster than $\tilde{\vec{x}}$ (as $N$ increases) for all values of $q$.\\

 

\vspace{-0.2cm}\noindent\textbf{Illustration on SSL.} The goal of SSL is to infer the label (or class) of all the nodes of a graph given a few pre-labelled nodes. Consider a partial labeling $\vec{Y}=\left(\vec{y}_1|\vec{y}_2|\ldots|\vec{y}_k\right)\in\mathbb{R}^{n\times k}$ of the nodes, where $k$ is the number of classes and $\vec{y}_l(i)$ is equal to $1$ if node $i$ is \emph{a priori} known to belong to class $l$ and $0$ otherwise. The objective is to find $k$ classification functions $\{\vec{f}_l\}_{l=1,\ldots,k}$ such that each $\vec{f}_l$ is on the one hand close to the labeling function $\vec{y}_l$ and on the other hand smooth on the graph, with a trade-off given by a parameter $\mu>0$. Depending on the choice of Laplacian used to define this graph smoothness (in the following, $\sigma=1$ corresponds to the combinatorial Laplacian, $\sigma=1/2$ to the normalized Laplacian, $\sigma=0$ to the random walk Laplacian), the explicit formulation of $\vec{f}_l$ can be written as (Prop. 2.2 of~\cite{avrachenkov_generalized_nodate}):
$
\vec{f}_l = \frac{\mu}{2+\mu}\left(\ma{I}-\frac{2}{2+\mu}\ma{D}^{-\sigma}\ma{W}\ma{D}^{\sigma-1}\right)^{-1}\vec{y}_l.
$
Note that this can be re-written, with $\ma{K}=\left(\ma{Q}+\ma{L}\right)^{-1}\ma{Q}$ and $\ma{Q} = \frac{\mu}{2}\ma{D}$, as:
\begin{align*}
\forall l=1,\ldots,k\qquad \vec{f}_l = \ma{D}^{1-\sigma}\ma{KD}^{\sigma-1}\vec{y}_l.
\end{align*}
Finally, once all classification functions $\vec{f}_l$ are computed, each node $i$ is classified 
in the class $\text{argmax}_{l}\; f_l(i)$. 

One may use our estimators to solve the SSL classification task: $\forall l$, i/~use the proposed estimators on the vector $\ma{D}^{\sigma-1}\vec{y}_l$ to estimate $\ma{KD}^{\sigma-1}\vec{y}_l$, ii/~left-multiply the result by $\ma{D}^{1-\sigma}$ and obtain an estimate of $\vec{f}_l$, iii/~once all functions $\vec{f}_l$ have been estimated, classify each node $i$ to $\text{argmax}_{l}\; f_l(i)$. In the following, we choose $\sigma=0$ and set $\mu=1$. 

We illustrate this on the Stochastic Block Model (SBM): a random graph model with communities. Consider a SBM with $n=3000$ nodes and two communities of equal size. We generate two scenarios: a well-structured (resp. fuzzy) setting with probability of intra-block connection $p_{\text{in}}=2\cdot10^{-2}$ (resp. $10^{-2}$) and inter-block connection $p_{\text{out}}=3\cdot10^{-3}$, which corresponds to a sparse graph with average degree $\simeq35$ (resp. $20$). The following experiment is performed: i/~choose $m$ nodes randomly in each community to serve as \emph{a priori} knowledge on the labels and use them to define the two label functions $\vec{y}_l$; ii/~compute the two classification functions $\vec{f}_l$ either via direct computation (method referred to as $\hat{\vec{x}}$) or via our proposed estimators; iii/~each node $i$ is classified in community $\text{argmax}\left[f_1(i), f_2(i)\right]$. The performance of the  community recovery is measured by the Adjusted Rand Index (ARI), a number between $-1$ and $1$: the larger it is, the more accurate the recovery of the ground truth. Results are shown in Fig.~\ref{fig:SSL_exp}. The estimator $\bar{\vec{x}}$ matches the performance of $\hat{\vec{x}}$ after $N=500$ forest realizations. Also, the smaller the amount of prior knowledge $m$ and the fuzzier the block structure, the harder it is to match $\hat{\vec{x}}$.  
A closer look at the sampled forests shows that some trees do not contain any labeled nodes, thus failing to propagate the label information. This proof-of-concept could be improved in various ways to avoid this difficulty --going beyond the scope of this paper.

\section{Conclusion}
\vspace{-0.1cm}
\label{sec:conclusion}
We provide an original and scalable method enabling to estimate graph smoothing operations in large graphs. In future work, we will further explore this deep link between RSFs and Laplacian-based numerical linear algebra, to apply these ideas to more advanced graph signal processing operations. 
\bibliographystyle{IEEEbib}
\bibliography{refs.bib}

\end{document}